\documentclass[conference,letter,10pt,final]{IEEEtran}
\usepackage[utf8]{inputenc}
\usepackage[T1]{fontenc}
\usepackage{graphicx}
\usepackage{grffile}
\usepackage{longtable}
\usepackage{wrapfig}
\usepackage{rotating}
\usepackage[normalem]{ulem}
\usepackage{amsmath}
\usepackage{textcomp}
\usepackage{amssymb}
\usepackage{capt-of}
\usepackage{hyperref}
\usepackage[utf8]{inputenc}
\usepackage[T1]{fontenc}
\usepackage{flushend}
\date{}
\title{Optimizing Cost per Click for Digital Advertising Campaigns}
\hypersetup{
 pdfauthor={Aditya Jain},
 pdftitle={Optimizing Cost per Click for Digital Advertising Campaigns},
 pdfkeywords={},
 pdfsubject={},
 pdfcreator={Emacs 27.1 (Org mode 9.5)}, 
 pdflang={British}}
\begin{document}

\makeatletter
\let\orgtitle\@title
\makeatother

\title{\orgtitle}

\author{
\IEEEauthorblockN{
   Aditya Jain\IEEEauthorrefmark{1},
   Sahil Khan\IEEEauthorrefmark{2}}

\IEEEauthorblockA{\IEEEauthorrefmark{1}
Data Scientist II R\&D, MiQ Digital }

\IEEEauthorblockA{\IEEEauthorrefmark{2}
Manager Data Science R\&D, MiQ Digital }
}

\maketitle

\begin{abstract}

Cost per click is a common metric to judge digital advertising campaign
performance. In this paper we discuss an approach that generates a feature
targeting recommendation to optimise cost per click. We also discuss a technique
to assign bid prices to features without compromising on the number of features
recommended.

Our approach utilises impression and click stream data sets corresponding to
real time auctions that we have won. The data contains information about device
type, website, RTB Exchange ID. We leverage data across all campaigns that we
have access to while ensuring that recommendations are sensitive to both
individual campaign level features and globally well performing features as
well. We model Bid recommendation around the hypothesis that a click is a
Bernoulli trial and click stream follows Binomial distribution which is then
updated based on live performance ensuring week over week improvement.

This approach has been live tested over 10 weeks across 5 campaigns. We see Cost
per click gains of 16-60\% and click through rate improvement of 42-137\%. At the
same time, the campaign delivery was competitive.

\end{abstract}

\section{Introduction}
\label{sec:orga9e14be}

For anyone accessing the internet, Digital marketing is not a new term as you
are targeted with advertisements on multiple platforms where businesses try to
reach the right audience interested in buying different products and services.
Digital marketing is an umbrella where all marketing channels like TV, Desktop
and mobile are used to reach out to a particular audience. Digital marketing
industry size in 2019 was around \$300-310 billions and is expected to grow in
2021 at 15-20\%. Digital marketing houses run primarily by agencies do look for
proven data science and machine learning methods to strike the right balance
between relevancy and growth. Online advertising is driven by the demand side
and supply side of the business primarily governed by real time bidding exchange
where real time auction takes place. Demand side is dominated by agencies and
clients who are looking for an audience interested in their products and
service. Publishers and sellers on the other hand dominate the Supply Side.
Supply side parties have a placeholder for advertisements to be shown on their
websites and mobile apps where traffic of relevant audience arrives and are thus
redirected to websites of clients ending up purchasing their products and
services.

Advertising on display and videos have been in the industry for long and people
do look for ways to innovate it. Although the industry is led by solution owners
or programmatic traders with their own assumptive intuition, Data Science
practices have opened doors for people to dig deep and dive in further to bring
the best value out of any click or conversion happening. With the algorithmic
approach, it has gone wide to help the media industry quickly. Targeting the
right audience for an advertisement does involve looking at a lot of factors and
that can be solved using a model based approach. One of the most common business
KPI that marketers look for when they want to call a managed campaign successful
is CTR - click through rate. Click Through Rate is a KPI which looks at ratio of
users who click the advertisement with respect to people shown an advertisement.
In marketing campaigns, you would observe a click through rate of 0.03\% while
through our methods, we tend to increase the performance of the campaigns by
1.5X to 2X making sure the other media related constraints are satisfied.
Technically a good click through rate depends on various factors including the
platform. A good CTR for adword's search page could be different from those of
Facebook's and would be largely different from media campaigns. It also normally
varies from one vertical to another, the place an advertisement is placed, size
of the advertisement creative and also the location where an advertisement is
targeted.

With this paper, we are looking at data driven techniques to optimise click
through rate. While looking at multi-level features selected for better clicks
based on initial exploration serves as an initial component, other important
aspects that we cover in this paper is the pipeline that allows us to scale our
solution to over a Billion rows.

Another form of scale relevant to this paper is scale of impressions. Getting
enough scale for any active campaign and at the same time making sure the cost
spent on getting those clicks while keeping the cost minimal is the problem we
are trying to solve in this paper.

\section{Data and Experimental Context}
\label{sec.context}
\subsection{Experimental Context}
\label{sec.framework}
Before we proceed with the discussion of our approach, the experimental context
and use-case of the final result needs to be addressed to better explain the
practical restrictions governing use of results obtained via the approach
outlined in this paper.

Many publications discuss approaches that can be utilised for modelling Click
Through Rate. Typical approach is modelling for \(P(x=1)\) using a classification
model. Neural Networks, SVMs, Decisions Trees, Random Forests, and various
boosted tree approaches have been show to work for this task. However, our
setting and requirement are different form these existing solutions.
Typical approaches usually are not very sensitive to the tail end of input
features. In practice, we have seen that for our campaigns the highest
performing features typically form the tail end of the distribution.
It is well known that optimising a model for tail end values is an uphill task.
Hence, we need an approach that optimises reasonably well for such features.

Our goal here is to recommend a feature combination along with a reasonable bid
value so that the aggregate cost per click is improved. Feature combination is a
set of contextual features that define where an impression can be delivered. For
example the following table shows a few valid feature combinations:

\begin{table}[htbp]
\caption{Example feature combinations}
\centering
\begin{tabular}{llrr}
Site Domain & Device Type & Size & Fold\\
\hline
analyticsindiamag.com & Mobile & 300x50 & 1\\
yahoo.com & Desktop & 300x250 & 0\\
\end{tabular}
\end{table}

Along with such feature combinations (sometimes referred to as context), we need
to send bid prices that we are willing to pay for an impression served at each
such context. We however have no way of specifying the exact number of
impressions that we wish to win for a particular feature combination. That
control is not available. Thus bid prices are the only other parameter that we
can control. Various methods exist that allow modelling of number of impressions
vs a bid price. However, all of them require auction level left censored data
that is not available to us \cite{cui2011}. Therefore, the approach discussed in
this paper focuses on assigning maximum bid price which is also the price at
which the expected CPC is equal to or lower than our target.

Digital advertising campaigns are very dynamic leading to varying week over week
performance of same feature sets. Any approach that is chosen for the task
should allow for constant feedback and iterative improvement. In case of
ill-performing feature set, the approach should be quick in updating its
recommendation.

At the same time the approach had to be compliant to GDPR, a European law
outlining privacy honouring requirements of data collection and processing.
Therefore, our approach does not use user level identifiers and operates at
aggregated feature level.

Our chosen approach fulfils all these requirements.

\subsection{Data}
\label{sec.data}
We use impression stream and click stream at the organisation level as the input
to our process. An impression stream data consists of all impressions that we
were able to serve at the account level. Similarly, click stream data consists
of every click that happened as a result of an advertisement shown by us.

Along with the information of an impression event or a click event, these
streams provide us information about the context of the ad impression. Typical
row from this data set contains information about the site domain where the ad
impression took place, time stamp of ad impression, device type, geographical
information like Zip code, Internet service provider etc. The complete data
dictionary contains well over 30 columns of which 7 columns that contain
information about price and targeting are of interest to us.

Three types of columns are present in our data set which convey
\begin{enumerate}
\item Context of ad slot
\item Cost of ad slot
\item Non Context information
\end{enumerate}

Context information indicates where the ad impression was shown and can be
directly used for targeting. This includes the following columns:
\begin{itemize}
\item \textbf{Timestamp}: Time stamp of click or impression
\item \textbf{Height}: Height of the image required by the ad slot
\item \textbf{Width}: Width of the image required by the ad slot
\item \textbf{Device Type}: Type of device Desktop, Mobile, or Tablet that this ad impression was shown
\item \textbf{Operating System}: Operating system of device
\item \textbf{Browser}: Browser type and version where this ad impression was shown
\item \textbf{Fold Position}: Above fold or below fold. Indicates if the advertisement is visible on page load or after scrolling down the page
\item \textbf{Geo Country}: Country where this ad impression was shown
\item \textbf{Geo Region}: DMA \cite{nielsen2000}
\item \textbf{Seller Member ID}: Seller via whom the inventory is made available
\item \textbf{Tag ID}: Unique ID of ad location on a website
\item \textbf{Publisher ID}: Unique ID of website owner
\item \textbf{Site Domain}: Mobile Application or Website
\end{itemize}

Non Context information includes following columns
\begin{itemize}
\item \textbf{Insertion Order ID}: Advertising campaign identifier
\item \textbf{Advertiser ID}: ID of Advertiser a particular impression or click belongs to
\item \textbf{Is Click}: 0 if not click, 1 if click
\end{itemize}

Cost information contains following columns:
\begin{itemize}
\item \textbf{Media Cost}: Cost of the impression in USD
\item \textbf{Data Cost}: Per impression cost of third party data used
\end{itemize}

Due to the targeting restrictions of our upstream provider, we combine Height
and Weight and create Size. Similarly, Geo country and Geo Region are combined
to form Geo targeting column. Actual realised cost of an ad impression is \(Media
Cost + Data Cost\). We use the aggregated amount for further analysis and
modelling.

From the click and impression stream, we prepare two data sets campaign level,
and network level with minor differences. Campaign level data contains all the
columns mentioned above that we filter from the larger data set. Network level
data however is not processed at campaign level. For this data set, we remove
the following columns:
\begin{itemize}
\item Insertion Order ID
\item Advertiser ID
\end{itemize}

The final data set contains approximately 1 Billion rows spanning a duration of
7 days. We will discuss more about the usage of these different data sets in
section \ref{sec.pipeline}.

\section{Modelling}
\label{sec.proposal}
An ad-impression can lead to two states that are relevant to this discussion. It
can either lead to a click or not lead to a click. We can thus say that a Click
is a binary random variable where the value \(0\) represents a non-click event and
\(1\) represents a click event. We are treating clicks, and impressions as independent events.

This allows us to model a click stream as a Bernoulli Trial \cite{papoulis2002,2020}
\subsection{Bernoulli Trial}
\label{sec:org24957f2}

\begin{itemize}
\item Let probability of a click be \(p_c\)
\item Then, Probability of no-click \(p_n = 1 - p_c\)
\item \(p_c + p_n = 1\)
\end{itemize}

Since we treat each impression as a Bernoulli Trial, it follows that a series of
such trials be modelled as a Binomial experiment where probability of getting
\(n\) clicks can be expressed as:
\begin{equation}
Pr(X = n) = { i \choose n } {p_c}^n(1 - p_c)^{i-n}
\end{equation}

From data, we can calculate the ratio of clicks vs total impressions. However,
consider a coin toss experiment. If we observe coin toss leading to 2 heads and
0 tails in two independent trails, does it follow that the coin only lands on
Heads?

This question leads us to the Beta Distribution.

\subsection{Beta Distribution}
\label{sec:org28f58dc}
Beta distribution is the conjugate prior for Binomial and Bernoulli
Distributions. Accordingly, we can write
\begin{equation}
f(p_c|n,n_i,i_i,i) \propto {p_c}^{n+n_i-1}(1-p_c)^{(i-n)+(i_i-n_i)-1} \label{eq.beta}
\end{equation}

where
\begin{itemize}
\item subscript \(i\) indicates imaginary trials
\end{itemize}

The expectation of \eqref{eq.beta} will give us the expected probability of click
\(P_c\) from click vs non click data.

For this we leverage Bayesian inference \cite{bishop2006} over Beta
Distribution as mentioned in equation \eqref{eq.beta_inf}

\begin{equation}
p(x=1|Data) = \frac{m+a}{m+a+l+b} \label{eq.beta_inf}
\end{equation}
where
\begin{itemize}
\item \(p(x=1)\) is the probability of Success
\item \(m\) is prior clicks
\item \(a\) is real clicks
\item \(l\) is prior non clicks
\item \(b\) is real non clicks
\end{itemize}

This affords us a very simple and explainable approach that we can use to
calculate expected click through rate or \(P_c\) from the data.

Per the Bayesian approach, we can use the same equation with updated data of \(a\)
and \(b\) to update our belief. This way, we can calculate the posterior
probability of clicks by simply adding new data to our data-set without
modifying any other part of the system.

\subsection{Final Approach}
\label{sec:org2ea2193}

Our final approach uses equation \eqref{eq.beta_inf} to calculate expected Cost as
well as expected Click through Rate. Along with this we use a heuristic measures
to prevent under delivery and high cost.

We utilise data from Network level feed as well as campaign level feed as
discussed in section \ref{sec.data}. The reasons for this are twofold:
\begin{itemize}
\item Prevent under delivery by using feature combinations with wider reach
extracted from network level data
\item Bootstrap performance of campaign from known high performing features from
network level data.
\end{itemize}

We first calculate network wide average impressions, and average number of
clicks for all feature combinations. This forms the prior part of equation
\eqref{eq.beta_inf}. For all feature combinations, we calculate adjusted click
through rate by adding the prior to their actual performance.

We repeat this step for cost column to give us a prior cost. Both these steps
allow us to handle feature combinations with few data points well.

The same steps are repeated for campaign level data where the prior is again
calculated at campaign level. Adjusted Cost and CTR are then calculated for all
campaign level features.

We then proceed with bid calculation targeting a specified CPC per the
logic below.
\begin{equation}
CTR = \frac{Click}{Impressions} \label{eq.ctr}
\end{equation}

\begin{equation}
CPC = \frac{Cost}{Click} \label{eq.cpc}
\end{equation}

\begin{equation}
adjusted\_ctr = \frac{prior\_click+click}{prior\_imp+imp} \label{eq.adj_ctr}
\end{equation}

\begin{equation}
\begin{aligned}
adjusted\_cost = prior\_cpm&*prior\_imp \\
&+ \\
 feature\_cpm&*feature\_imp
\end{aligned}
\end{equation}
\begin{equation}
adjusted\_imp = prior\_imp + feature\_imp
\end{equation}
\begin{equation}
adjusted\_cpm = \frac{adjusted\_cost}{adjusted\_imp}*1000
\label{eq.adj_cpm}
\end{equation}

\begin{equation}
CPM = CPC*CTR*1000 \label{eq.cpm_ctr}
\end{equation}
Substituting \(CPC\) in equation \eqref{eq.cpm_ctr} with a known target, we can
calculate the max affordable \(CPM\). By substituting \(CTR\) with \(adjusted\_ctr\)
in equation \eqref{eq.cpm_ctr} we can calculate the highest bid price we can
recommend given the expected CTR. In this step we introduce a parameter
\(optimization\_fraction\). Since the goal of this approach is to optimise \(CPC\),
we multiply this variable with the obtained \(CPM\) before recommending it to the
users.
This enables us to always push recommendations that would perform better than
the rest of the campaign in terms of \(CPC\).
Very aggressive value of \(optimization\_fraction\) leads to severe under
delivery. Hence, it is advisable to test a few variations or modify this value
automatically in a feedback loop.

\section{Implementation Overview}
\label{sec.pipeline}
\begin{figure*}
\centering
\includegraphics[width=.9\linewidth]{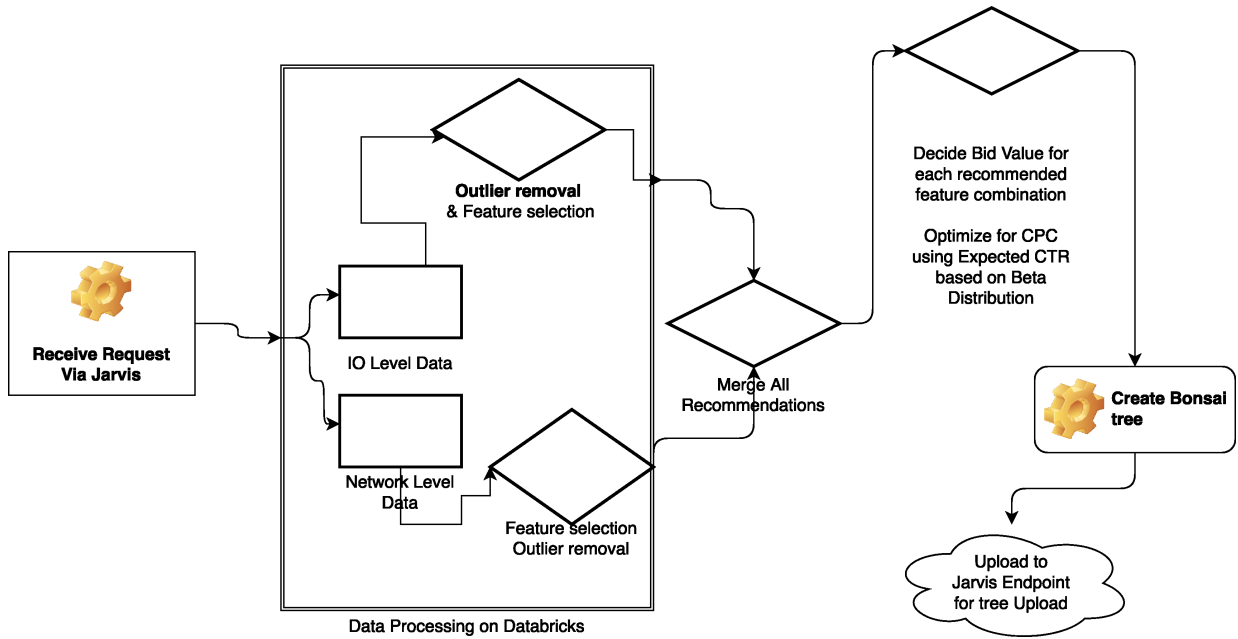}
\caption{High level pipeline architecture \label{img.pipeline}}
\end{figure*}

Any task in Digital Advertising industry has to handle at least a few terabytes
of data. The approach in this paper is no different and needs to scale to \textasciitilde{}24TB
of raw input data. PySpark or in our case Databricks is the go to platform.

Figure \ref{img.pipeline} outlines the overall design of pipeline that  we are
currently using to generate recommendations. It is divided into 3 parts
\begin{itemize}
\item Request creation
\item Aggregation and Generation of recommendation
\item Activation
\end{itemize}

Jarvis is an internal tool that takes care of receiving requests for
recommendation which is then processed in batch mode once or twice per week.

The bulk of processing happens on Databricks. The first step is to load the raw
feeds from S3. For the purpose of these experiments, we loaded 7 days of
impression and click stream feeds. Following are the steps that are performed on
network level data:
\begin{itemize}
\item Filter for relevant Geographical region.
\item Group by data with relevant fields
\item Remove outliers outside 2 standard deviation
\item Calculate average impression \& click for use as prior
\item Add prior to all feature combinations generated
\item Sort by adjusted CTR
\item Filter the feed for top 100K impressions with the highest adjusted CTR
\item These features are common for all campaigns however bid values are different
across each campaign
\end{itemize}

We perform similar step on campaign data. Prior impressions and clicks are
calculated at per campaign level. Another difference is that we do not filter
campaign data for top 100K impressions. All impressions and feature combinations
are used albeit at lower bids.
Once both feeds are individually processed, we proceed with a merger of
recommendations from both these sources based on the requested scale of
recommendation. Typically, 30\% of scale is served from network level features.

Next step is to calculate the bid values using methods discussed in section \ref{sec.proposal}.
Subsequent steps involve packaging these results into required format and
uploading them to our upstream service provider.

We repeat this process twice every week to ensure that bad performing
features are kept in check.

\subsection{Effectiveness of Feedback loop}
\label{sec:org3359519}

Our hypothesis is that every feature that is not optimally performing will
eventually face reduced bids till it starts performing better.

Let's assume a feature combination \(T\) that has only delivered 100 impressions
so far and has received exactly 1 click. At this point there  isn't enough
information about \(T\) to allow us to make an informed decision. Therefore, we
add the prior values calculated from campaign data.

For the sake of argument let's assume that the prior values are at 1 click and
1000 impression. After adding this to the data of \(T\), the effective \(CTR\) now
becomes

\(adj\_ctr = \frac{1+1}{100+1000} = 0.18\%\)

This adjusted CTR value is very high and consequently \(T\) receives a very high
bid value.

In the next iteration of the pipeline, there are 3 possible cases
\begin{enumerate}
\item \(T\) is performing really well
\item \(T\) is delivering a lot of impressions and thus costing us a lot without
getting us a lot of clicks
\item \(T\) is not able to deliver at all. The delivery is stuck at 100 impressions.
\end{enumerate}

Case 3 is trivial. The algorithm will arrive as bids same as the last time, and
we will not see a lot of delivery again in the next week. This is OK as long as
other features are delivering sufficient inventory.

Case 1 is also trivial. The algorithm will recalculate the adjusted CTR and
increase the bids as applicable.

Case 2 is where we need to ensure that bad performing features stop delivering
or deliver at a lower cost per thousand impressions thereby increasing the
effective CPC. Let's assume that the total impressions delivered by \(T\) in this
case is 10,000 without any new clicks.

By calculating the adjusted ctr, we get

\(adj\_ctr = \frac{1}{11000} = 0.009\%\)

This time around, the algorithm will reduce the bid value allocated to feature
combination \(T\) as governed by equation \eqref{eq.cpm_ctr}. Since adjusted CTR is
in the numerator of this equation, the effective bids allocated to \(T\) will be
very low as required by the equally bad performance.

Hence, if a high bid value is assigned to a bad performing feature yet unknown
to us, it is benign if it falls under case 3. If it falls under Case 2, we can
be sure that the bids will be reduced in response. This dynamic nature of our
approach make it responsive to bad performing features and ensures that campaign
budget is not wasted.

\section{Tests and Results}
\label{sec:org73484a1}

\begin{table*}[htbp]
\caption{Delivery comparison of Existing lines vs new recommended line \label{tab-performance-imp}}
\centering
\begin{tabular}{rrllrrl}
\hline
Impressions & Daily Budget & Delivery \% & \textbf{Campaign} & Impressions & LI Budget Daily & Delivery \%\\
\hline
\textbf{C} & \textbf{C} & \textbf{C} & \(\leftarrow\) Type \(\rightarrow\) & \textbf{R} & \textbf{R} & \textbf{R}\\
\hline
4626434 & 16170000 & 28.61\% & \textbf{A} & 817478 & 2560000 & 31.93\%\\
6537127 & 13914000 & 46.98\% & \textbf{B} & 1124843 & 960000 & 117.17\%\\
8077947 & 57446000 & 14.06\% & \textbf{C} & 1201979 & 2860000 & 42.03\%\\
3161197 & 6910000 & 45.75\% & \textbf{D} & 170969 & 9600000 & 17.81\%\\
2965780 & N/A & N/A & \textbf{E} & 104277 & N/A & N/A\\
\hline
\end{tabular}
\end{table*}

\begin{table*}[htbp]
\caption{KPI comparison of Existing lines vs new recommended line \label{tab-performance-kpi}}
\centering
\begin{tabular}{rrrrllrrrrl}
\hline
Clicks & Media Cost & CPC & CPM & CTR & \textbf{Campaign} & Clicks & Media Cost & CPC & CPM & CTR\\
\hline
\textbf{C} & \textbf{C} & \textbf{C} & \textbf{C} & \textbf{C} & \(\leftarrow\) Type  \(\rightarrow\) & \textbf{R} & \textbf{R} & \textbf{R} & \textbf{R} & \textbf{R}\\
\hline
3900 & 8080 & 2.07 & 1.75 & 0.08\% & \textbf{A} & 1538 & 1693.233 & 1.1 & 2.07 & 0.19\%\\
3616 & 12486 & 3.45 & 1.91 & 0.06\% & \textbf{B} & 1019 & 1378.6872 & 1.35 & 1.23 & 0.09\%\\
5823 & 11472 & 1.42 & 1.97 & 0.07\% & \textbf{C} & 1204 & 1435.88 & 1.19 & 1.19 & 0.10\%\\
1882 & 6610 & 3.51 & 2.09 & 0.06\% & \textbf{D} & 336 & 321 & 0.96 & 1.88 & 0.20\%\\
5594 & 6626 & 1.42 & 1.97 & 0.07\% & \textbf{E} & 472 & 258 & 0.56 & 2.48 & 0.45\%\\
\hline
\end{tabular}
\end{table*}

We tested our approach on 5 live campaigns in the US Region across different
verticals and campaign configuration for a duration starting from Late September
to Early December. Within each of these campaigns, a new strategy (Line Item)
was created and associated with our recommendations. Other strategies that were
already delivering on these campaigns included ones optimising for Impressions,
CPC, Viewability. No change was made to other line items of these test
campaigns.

Tables \ref{tab-performance-imp} and \ref{tab-performance-kpi} compare the result of
existing strategies indicated by \textbf{C} against type with recommended strategies
indicated by \textbf{R} against type. Campaign \textbf{A} to \textbf{C} had a greater geographical
coverage. Campaign \textbf{D} was configured to deliver on a very restricted
geographical area akin to a district. Campaign \textbf{E} was a geo-fence campaign
using 3\textsuperscript{rd} party data.

In table \ref{tab-performance-imp}, the Impressions column indicates the total
number of advertisements show during the test period for a campaign. The daily
budget column contains the sum of individual targets of each strategy.
Typically, stakeholders over-allocate strategies to ensure campaign delivery.
Therefore, we see the Delivery\% column containing numbers much below 50\%. In
practice, the Delivery\% of our recommendations should be comparable to
corresponding existing strategies.

On the delivery front, we see that Campaigns A to C have a much higher delivery
percentage for our recommendations. This percentage when higher indicates that
we are able to deliver more than our fair share of the impressions. In campaign
D our recommendation as only able to reach 17.81\% delivery whereas existing
strategies delivered 45.75\%. We attribute this to the strict geographic
requirement of the campaign.

On the KPI front in table \ref{tab-performance-kpi} we see that campaigns that
were able to fulfil delivery requirements also have 42.8\% to 137.5\% better Click
through Rate(CTR) and at the same time have 16.19\% to 60.86\% better Cost Per
Click(CPC).

CPM across all well delivering campaigns is lower except for campaign A where it
is 18.28\% higher than the corresponding strategies. However, this is more than
made up by the much better CTR allowing the line to achieve a lower CPC with
respect to control lines.

Campaigns D and E under delivered and the corresponding delivery is much lower
than required. However, even in such cases our approach resulted in  much lower CPC
and much higher CTR. Campaign D realised 233\% improvement in CTR corresponding
to 72\% reduction in CPC. Campaign E realised 641\% improvement in CTR
corresponding to 60\% reduction in CPC.

Thus far, our approach has been able to meet the primary goal of improving Cost
per Click for each of the campaigns. After monitoring the week over week
performance of these campaigns for the test duration, we can say
that the approach is able to react quickly to performance changes, thus
satisfying our requirement of responsiveness.

\section{Conclusion and Future Work}
\label{sec:orgb429bb0}
In this paper we have discussed the effectiveness of modelling a Click event as a Bernoulli Trial.
In digital advertising, many events like Converts, Views, Video completion are
suitable candidates for application of this approach. We have seen certain edge
cases like restrictive geographical targeting that have resulted in low
impression delivery. We would like to explore variations to this approach that
will enable us to guarantee delivery for such campaigns. A reinforcement
learning approach to modify the parameters of our approach will reduce
the manual intervention required in cases of extreme delivery.

\bibliographystyle{IEEEtran}
\bibliography{refs}

\begin{thebibliography}{1}
\providecommand{\url}[1]{#1}
\csname url@samestyle\endcsname
\providecommand{\newblock}{\relax}
\providecommand{\bibinfo}[2]{#2}
\providecommand{\BIBentrySTDinterwordspacing}{\spaceskip=0pt\relax}
\providecommand{\BIBentryALTinterwordstretchfactor}{4}
\providecommand{\BIBentryALTinterwordspacing}{\spaceskip=\fontdimen2\font plus
\BIBentryALTinterwordstretchfactor\fontdimen3\font minus
  \fontdimen4\font\relax}
\providecommand{\BIBforeignlanguage}[2]{{%
\expandafter\ifx\csname l@#1\endcsname\relax
\typeout{** WARNING: IEEEtran.bst: No hyphenation pattern has been}%
\typeout{** loaded for the language `#1'. Using the pattern for}%
\typeout{** the default language instead.}%
\else
\language=\csname l@#1\endcsname
\fi
#2}}
\providecommand{\BIBdecl}{\relax}
\BIBdecl

\bibitem{cui2011}
Y.~Cui, R.~Zhang, W.~Li, and J.~Mao, ``Bid landscape forecasting in online ad
  exchange marketplace,'' in \emph{Proceedings of the 17th {{ACM SIGKDD}}
  International Conference on {{Knowledge}} Discovery and Data Mining}, 2011,
  pp. 265--273.

\bibitem{nielsen2000}
A.~C. Nielsen, ``{{US Designated Market Areas}},'' 2000.

\bibitem{papoulis2002}
A.~Papoulis and S.~U. Pillai, \emph{Probability, Random Variables, and
  Stochastic Processes}.\hskip 1em plus 0.5em minus 0.4em\relax {Tata
  McGraw-Hill Education}, 2002.

\bibitem{2020}
``\BIBforeignlanguage{en}{Bernoulli trial},''
  \emph{\BIBforeignlanguage{en}{Wikipedia}}, Oct. 2020.

\bibitem{bishop2006}
C.~M. Bishop, \emph{Pattern Recognition and Machine Learning}.\hskip 1em plus
  0.5em minus 0.4em\relax {springer}, 2006.

\end{thebibliography}
\end{document}